\documentclass[a4paper,fleqn,usenatbib]{mnras}

\usepackage[T1]{fontenc}
\usepackage{ae,aecompl}
\usepackage[normalem]{ulem}

\usepackage{graphicx}	
\usepackage{amsmath}	
\usepackage{amssymb}	

\title[Spin-period variations in the intermediate polar RX J2133.7+5107]{Spin-period variations in the intermediate polar RX J2133.7+5107}

\author[V. V. Breus et al.]{
V. Breus$^{1},$\thanks{E-mail: vitaly.breus@gmail.com (VB)}
I. L. Andronov$^{1}$,
P. Dubovsky$^{2}$,
Y. Kim$^{3,4}$,
J. N. Yoon$^{3,4}$,
K. Petr\'ik$^{5}$
\\
$^{1}$Department of Mathematics, Physics and Astronomy, Odesa National Maritime University, Mechnikova 34, UA-65029 Odesa, Ukraine\\
$^{2}$Vihorlat Astronomical Observatory, Mierova 4, SK-06601 Humenne, Slovak Republic\\
$^{3}$Department of astronomy and space science, Chungbuk National University, 361-763, Cheongju, Korea \\
$^{4}$Institute for Basic Science Research, Chungbuk National University, 361-763, Cheongju, Korea \\
$^{5}$M. R. Stefanik Observatory and Planetarium, Sladkovicova 41, SK-92001 Hlohovec, Slovak Republic\\
}

\date{Accepted 2024 XXX XX. Received 2024 XXX XX; in original form 2024 XXX XX}

\pubyear{2024}

\begin{document}
\label{firstpage}
\pagerange{\pageref{firstpage}--\pageref{lastpage}}
\maketitle

\begin{abstract}

We report the results of long-term time series photometry on RX J2133.7+5107 (also known as 1RXS J213344.1+510725) obtained at several observatories. Using data taken during 17 years, we determined the current value of the spin period of $570.811470$ seconds with the formal accuracy of $0.000006$ seconds and a spin-up of the white dwarf with a characteristic time of $1.483(1)\times10^5$ years. This is even faster than that reported previously and, if confirmed, makes this object have one of the fastest spin-up timescales of all known intermediate polars. We derived an improved value of the superhump period of the system to be $0^d.280130(1)$. Superhump maxima timings are moving on the phase curve from season to season, showing non-monotonic changes, without a change in superhump period.

\end{abstract}




\section{Introduction}
Cataclysmic variables are close binary star systems in which a white dwarf accretes mass from a low-mass secondary star via Roche-lobe overflow. Intermediate polars, often called IPs or DQ Her type stars, consist of a magnetic white dwarf and a lower main-sequence star. Due to the high angular momentum of the plasma leaving the inner Lagrangian point, the stream can not be accreted directly by the compact star, and instead it forms an accretion disk around the white dwarf. The most comprehensive review was presented by \citet{Warner1995}.

Usually, intermediate polars show two types of optical variability. The orbital variability is caused by the rotation of the binary, while the spin variability is caused by the rotation of the white dwarf with one or two accretion columns on its surface. Therefore, the light curve is frequently a superposition of two different periodic variations as well as aperiodic processes like flickering or outbursts. Because the accreting matter transfers mass and angular momentum to the white dwarf, the white dwarf spin undergoes period variations, which may be studied photometrically (\citet{2011JKASRXJ1803}).
Some systems show spin period variations, which are caused by varying mass transfer of the system and by varying orientation of the magnetic axis and/or rotational axis (precession) of the white dwarf. Some intermediate polars show "spin-orbital" beat periods (see the review by \citet{Patterson1994}). Some objects show unusual effects, like periodic modulation of the spin phases with the orbital phase (see \citet{Caosp2020mu}).

The X-ray source RX J2133.7+5107 was extracted from the ROSAT Galactic Plane Survey and classified as cataclysmic variable by \citet{Motch1998}.
\citet{Bonnet2006} classified it as an intermediate polar, determined the periods $P_{\omega} = 570.823 \pm 0.013$ s and $P_{\Omega} = 7.193 \pm 0.016$ hr and interpreted as spin and orbital periods respectively based on optical photometry and spectroscopy. \citet{DeMartino2006} determined the orbital period of the system $P_{\Omega} = 9.87 \pm 0.03$ hr, and the white dwarf spin period, $P_{\omega} = 693.01 \pm 0.06$ s, from medium resolution optical spectroscopy.

\citet{Katajainen2007} discovered circularly polarized emission in all UBVRI bands with the level of polarization up to 3\%. \citet{Anzolin2009} studied broad-band properties of this object using XMM-Newton and Suzaku data and confirmed as members of the intermediate polars class.

\citet{Thorstensen2010} improved the accuracy of previously published orbital period of RX J2133.7+5107 of $P_{\Omega}0^d.297431(5)$ with initial epoch HJD $2453074.033(3)$.

\citet{demiguel2017} observed this object in 2010-2016. They observed a modulation with 6.72-h period, interpreted as a negative superhump and also obtained a parabolic fit to the (O-C) diagram of the spin maxima that corresponds to $P_{\omega}/|\dot{P_{\omega}|} = 0.17\times10^6$ yr.

\citet{Covington2022} presented optical photometry of six intermediate polars that exhibit transitions to a low-flux state, and found that for RX J2133.7+5107 the temporal properties do not change in the low state. They mention RX J2133.7+5107 is unusual in that the low states only lasted a matter of days, and the only change in the timing signatures came when the optical flux was slightly higher than typical.

The paper is organized as follows: in the next section we describe the photometric 
data gathered for RX J2133.7+5107 and the instruments used, Section~\ref{sec:SpinPeriod} presents the methods used for derivation of the high precision period values. In Sections~\ref{sec:PeriodAnalysis} and \ref{sec:LongTerm} we describe the results obtained for the orbital and spin periods detected in the light curves of RX J2133.7+5107. We discuss our findings and conclsions in the Section~\ref{sec:Conclusions}.


\section{Observations and data reduction}
\label{sec:Observations}

An observing log is listed in Table~\ref{tabobs}, the baseline is presented in days.

The largest set of photometric data used in this research was obtained with the 60~cm Zeiss Cassegrain telescope at the M. R. Stefanik Observatory and Planetarium in Hlohovec, Slovakia (ZC600). At this site two instruments were used: 2007-2014 the SBIG ST-9 camera, while the ATIK 383L CCD was attached to this telescope in 2015-2024.

A second set of data was obtained using the 1~m Vihorlat National Telescope at the Astronomical Observatory on Kolonica Saddle, Slovakia (VNT). In 2007 the SBIG ST9-XE camera was used, in 2008 - SBIG ST8, while the FLI PL1001E was attached to this telescope in 2009-2024.

In 2014-2016 we gathered photometric observations with the the 40~cm Maksutow telescope at the Astronomical Observatory of the Jagiellonian University in Krakow, Poland (OAUJM), equipped with SBIG ST-7 camera. 1 night was obtained using 50~cm Cassegrain of the same observatory with Andor DZ936 BV camera (OAUJ50).

In 2005 - 2013 we acquired photometric observations using a 2K x 2K CCD camera mounted on the LOAO 1.0m telescope, Arizona (LOAO).

Additionally, our analysis includes 1 night obtained with 20~cm Meade LX200 at the M. R. Stefanik Observatory and Planetarium in Hlohovec, Slovakia, equipped with SBIG ST-9 in 2012 and 7 nights using Starlight Xpress SX-694 camera in 2019 (Meade). 7 runs were obtained using 50-cm reflector of Baja astronomical observatory, Baja, Hungary (50CM) using Apogee camera and 1 run using Celestron 14" at the Kolonica Observatory, equipped with Moravian Instruments G2-1600 CCD camera (C14).

Similarly to the data used by \citet{demiguel2017}, we have both long time series and shorter runs obtained for the purpose of tracking the spin period variations. Most of our data were taken in R filter or with alternating wide band V and R filters. Some observational runs were taken unfiltered. Exposure times were between 30--60 seconds, adjusted according to instrument and filters used and weather conditions. In total, RX J2133.7+5107 was observed on more than 208 nights, totalling more than 283 time series in different filters during 2007-2023.

The reduction, consisting of calibration of scientific images for bias, dark and flat-field and extraction of instrumental magnitudes, was carried out with the MUNIWIN software \citep{Munipack} or CoLiTecVS (\citet{colitec1}, \citet{colitec2}). We used nearby comparison stars, listed in the AAVSO Variable Star Plotter\footnote{https://www.aavso.org/apps/vsp/} database. The comparison stars were checked to be constant in our entire data set. The final derivation of magnitudes was obtained using the multiple comparison stars method described by \citet{KimAndMC2004} and implemented in Multi-Column View by \citet{AndBak2004} (hereafter named MCV). Barycentric corrections was applied to all geocentric Julian dates.

\begin{table}
	\centering
	\caption{The log of observations}
	\label{tabobs}
\begin{tabular}{cccccc}
\hline
Year & Instrument & BJD range & Baseline & Runs & Nights \\
\hline
2005 & LOAO & 53671 - 53671 & 1 & 1 & 1 \\
2007 & ZC600 & 54302 - 54321 & 19 & 12 & 6 \\
2007 & VNT & 54410 - 54433 & 23 & 2 & 2 \\
2008 & ZC600 & 54683 - 54690 & 7 & 10 & 5 \\
2008 & VNT & 54689 - 54774 & 85 & 5 & 5 \\
2009 & VNT & 55014 - 55101 & 87 & 6 & 6 \\
2009 & 50CM & 55054 - 55074 & 20 & 10 & 5 \\
2010 & VNT & 55352 - 55480 & 128 & 6 & 6 \\
2010 & LOAO & 55450 - 55454 & 4 & 2 & 2 \\
2011 & VNT & 55712 - 55877 & 165 & 6 & 6 \\
2011 & 50CM & 55834 - 55837 & 3 & 4 & 2 \\
2011 & LOAO & 55845 - 55850 & 5 & 2 & 2 \\
2012 & VNT & 56075 - 56247 & 172 & 9 & 8 \\
2012 & LOAO & 56085 - 56207 & 122 & 6 & 6 \\
2012 & Meade & 56097 - 56097 & 1 & 2 & 1 \\
2012 & ZC600 & 56098 - 56132 & 34 & 10 & 6 \\
2012 & C14 & 56105 - 56105 & 1 & 2 & 1 \\
2013 & LOAO & 56443 - 56448 & 5 & 3 & 3 \\
2013 & VNT & 56456 - 56589 & 133 & 10 & 10 \\
2013 & ZC600 & 56484 - 56511 & 27 & 11 & 6 \\
2014 & OAUJ50 & 56798 - 56798 & 1 & 2 & 1 \\
2014 & OAUJM & 56799 - 56800 & 1 & 4 & 2 \\
2014 & ZC600 & 56817 - 56827 & 10 & 9 & 5 \\
2014 & VNT & 56825 - 56972 & 147 & 5 & 5 \\
2015 & OAUJM & 57158 - 57172 & 14 & 8 & 4 \\
2015 & ZC600 & 57181 - 57214 & 33 & 10 & 10 \\
2015 & VNT & 57183 - 57326 & 143 & 3 & 3 \\
2016 & VNT & 57563 - 57716 & 153 & 5 & 5 \\
2016 & OAUJM & 57565 - 57580 & 15 & 10 & 5 \\
2016 & ZC600 & 57588 - 57611 & 23 & 8 & 8 \\
2017 & VNT & 57901 - 57908 & 7 & 2 & 2 \\
2017 & ZC600 & 57915 - 57969 & 54 & 15 & 15 \\
2018 & ZC600 & 58274 - 58381 & 107 & 41 & 22 \\
2018 & VNT & 58281 - 58429 & 148 & 3 & 3 \\
2019 & VNT & 58629 - 58724 & 95 & 4 & 4 \\
2019 & Meade & 58655 - 58680 & 25 & 7 & 7 \\
2019 & ZC600 & 58717 - 58728 & 11 & 12 & 6 \\
2020 & VNT & 59072 - 59111 & 39 & 3 & 3 \\
2021 & ZC600 & 59369 - 59369 & 1 & 1 & 1 \\
2021 & VNT & 59369 - 59518 & 149 & 10 & 8 \\
2022 & VNT & 59745 - 59932 & 187 & 4 & 4 \\
2022 & ZC600 & 59781 - 59846 & 65 & 6 & 4 \\
2023 & ZC600 & 60122 - 60197 & 75 & 5 & 5 \\
2023 & VNT & 60192 - 60192 & 1 & 1 & 1 \\
2024 & ZC600 & 60485 - 60485 & 1 & 1 & 1 \\
\hline
\end{tabular}
\end{table}


\section{Variation of the spin period}
\label{sec:SpinPeriod}
To determine spin and superhump maxima timings we used the trigonometric polynomial approximation of the light curve implemented in MCV applied for each filter separately. We choose a 2-periodic model for smoothing as described by \cite{breus2019wga} and similar to the method used by \cite{exhya2013}. This way we obtain one spin maxima timing per night per filter.
Along with our own data we also determined timings from all long data sets available in the AAVSO database and included in the analysis all spin timings published by \citet{demiguel2017} in supplementary data.

We used an initial value of the spin period that corresponds to a frequency of 151.35991 cycles $d^{-1}$ and an initial epoch of $2455461.39664$ \citep{demiguel2017} to calculate the (O-C) diagram shown in Fig.~\ref{fig:fig_spinoc} (a table is available in appendix~\ref{sec:online}).

We performed computations to derive best fits to the (O-C) diagram. A parabolic fit yields 
\begin{equation}
\begin{split}
\phi= -1.818(3) - 23.46(1)\times10^{-6}~\Delta~E\\- 6.097(5)\times10^{-11}~(\Delta~E)^2
\end{split}
\end{equation}
Here, $\Delta~E = E - E_0$, and $E_0=350928$ is an integer cycle number close to mean time of observations. This fit may correspond to the current value of the spin period of $0^d.00660661424\pm0.00000000007$, $\dot{P}=-1.219\times10^{-10}$. It suggests that the spin period decreases on a characteristic time-scale $P_{\omega}/|\dot{P_{\omega}|} = 1.483(1)\times10^5$ yr.

We consider possible fits with higher orders. A cubic fit has the coefficient $Q_3$ about $25\sigma$, but forth order polynomial has statistically significant coefficient $Q_4$ about $10\sigma$ and statistically better fits the data. These fits are
\begin{equation}
\begin{split}
\phi= -1.792(3) - 2.318(1)\times10^{-5}~\Delta~E\\
- 6.20(1)\times10^{-11}~(\Delta~E)^2 - 2.34(17)\times10^{-18}~(\Delta~E)^3\\ + 5.2(5)\times10^{-24}~(\Delta~E)^4
\end{split}
\end{equation}

and 

\begin{equation}
\begin{split}
\phi= -1.809(3) - 2.310(2)\times10^{-5}~\Delta~E\\
- 6.116(4)\times10^{-11}~(\Delta~E)^2 - 3.35(13)\times10^{-18}~(\Delta~E)^3
\end{split}
\end{equation}

\begin{figure}
	\includegraphics[width=\columnwidth,trim={0.0cm 0 0.5cm 0.0cm},clip]{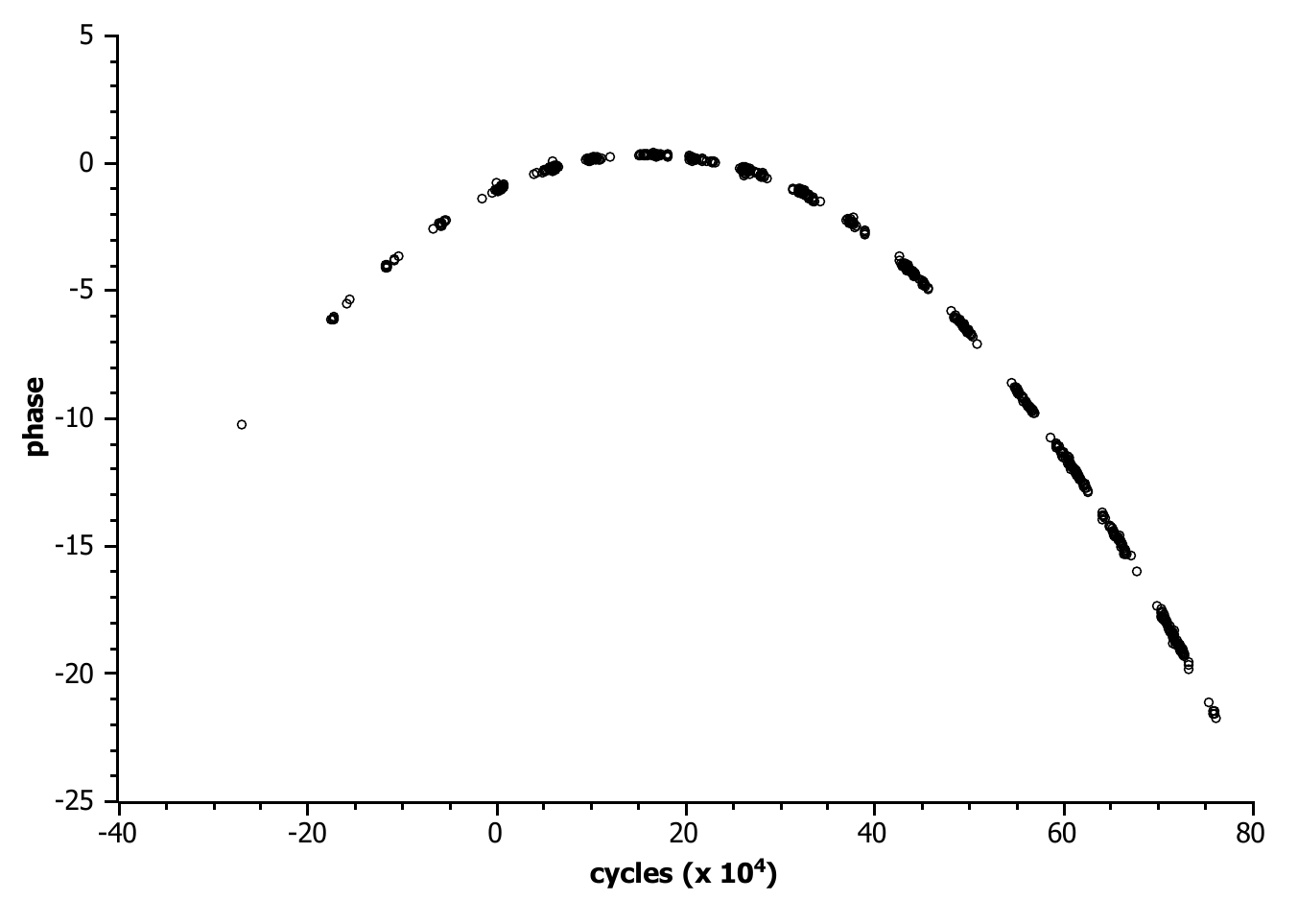}
    \caption{The dependence of phase of spin maxima on the cycle number (all available data) is shown.}
    \label{fig:fig_spinoc}
\end{figure}

\begin{figure}
	\includegraphics[width=\columnwidth,trim={0.0cm 0 0.5cm 0.0cm},clip]{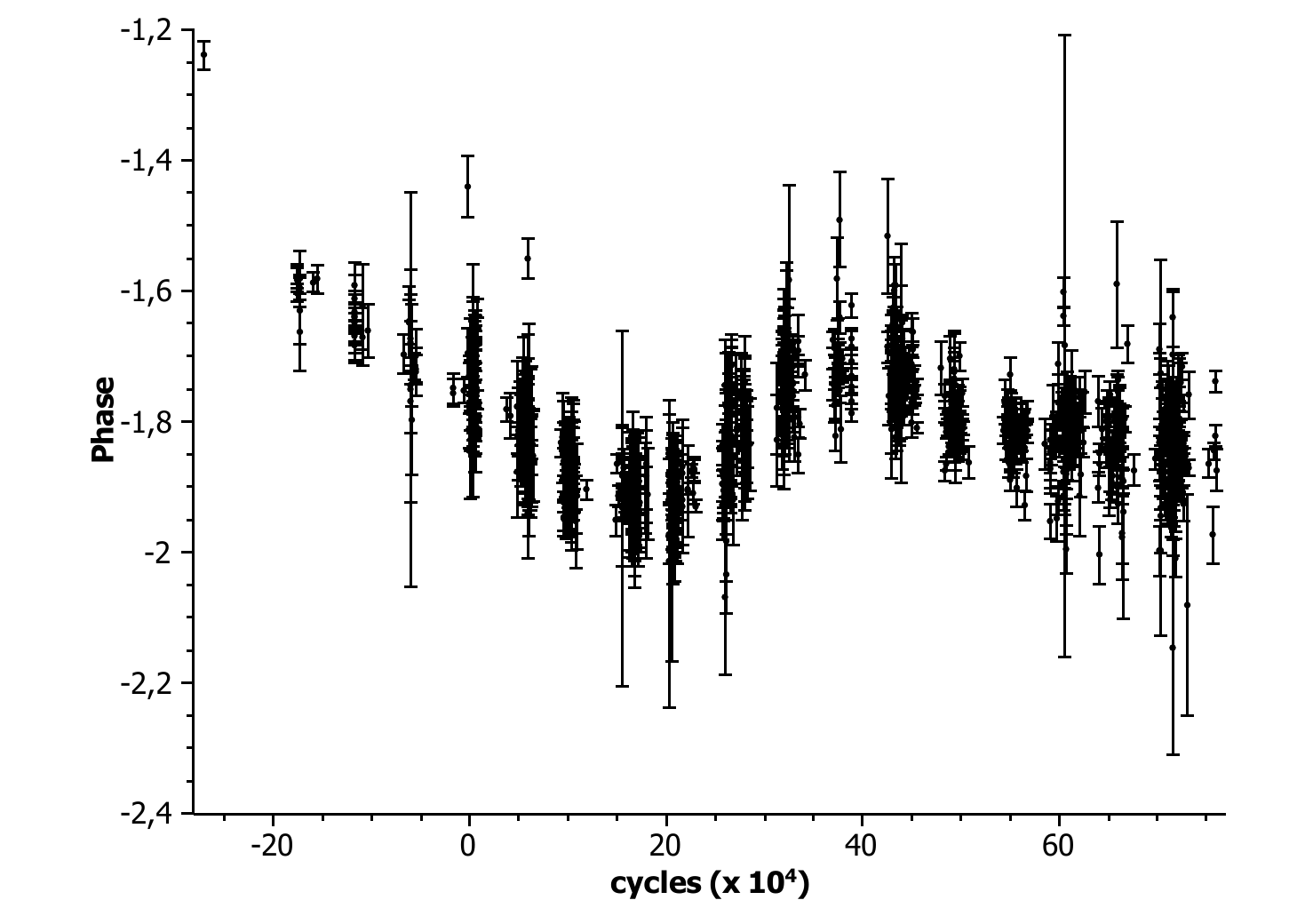}
    \caption{The residuals from the parabolic fit are shown.}
    \label{fig:fig_spinocresi}
\end{figure}


\section{Period Analysis}
\label{sec:PeriodAnalysis}

\begin{figure}
	\includegraphics[width=\columnwidth,trim={0.0cm 0 0.5cm 0.0cm},clip]{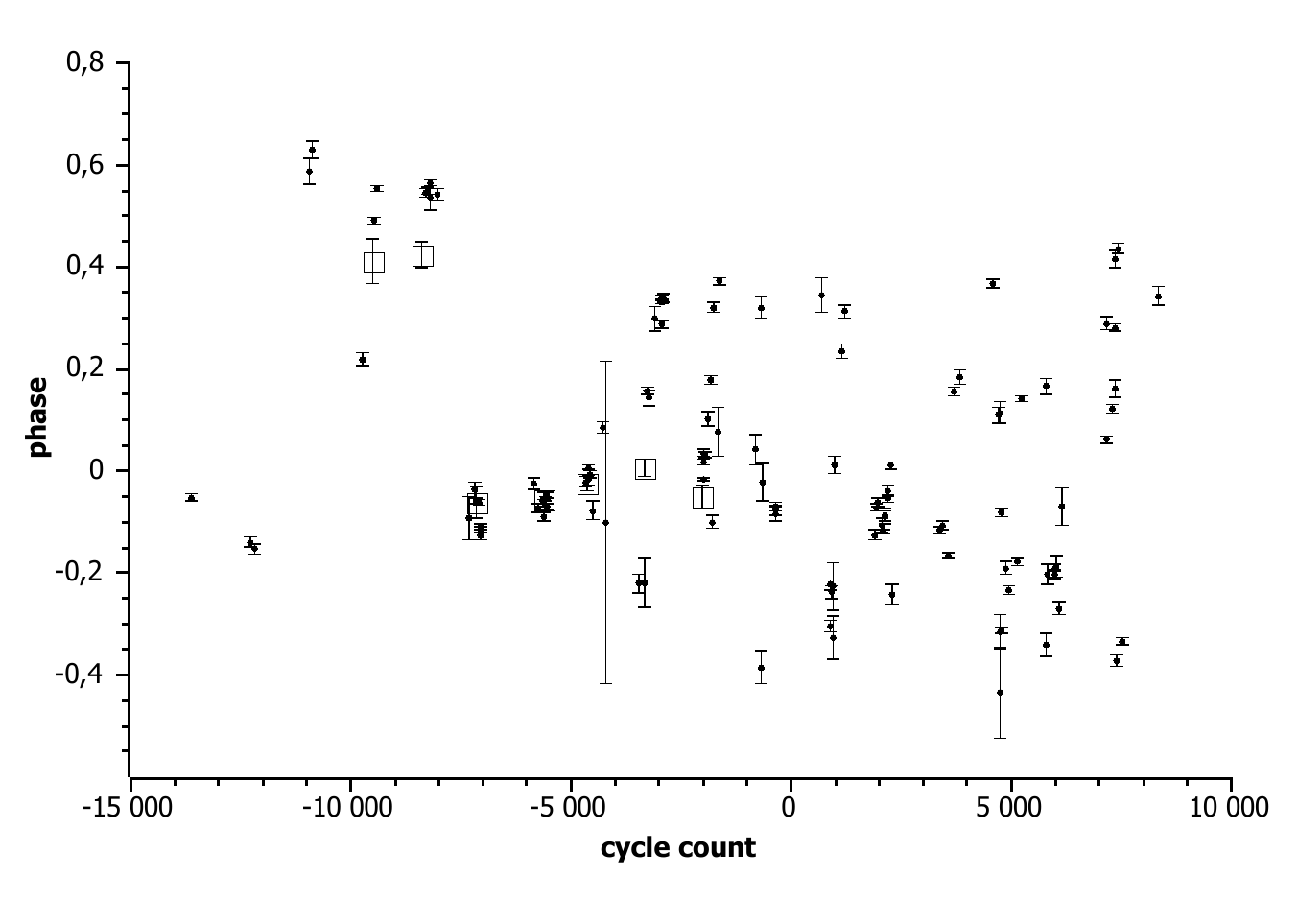}
    \caption{The dependence of phases of superhump maxima on the cycle number. This paper (black dots) and seasonal maxima timings published by \citet{demiguel2017} (open rectangles)}
    \label{fig:fig_ocsup}
\end{figure}

We formed seasonal light curves - one per each year, separately for each filter and used instrument, to avoid transforming instrumental magnitudes into a standard magnitude system. For some seasons and instruments with largest time intervals it was split into two light curves.

We may conclude that there is significant phase shift from season to season especially visible in Fig.~5. published by \citet{demiguel2017} (for the points for 2012 - 2016), but at the same time points for 2010 and 2011 are shifted for 0.5 periods. That means the test frequency should be improved and the variations in the periodicity is non-monotonic.

Using the method mentioned in Section~\ref{sec:SpinPeriod} we determined superhump maxima timings in two iterations. From the first iteration we obtained superhump variability period of $0^d.280130(1)$. Timings for our observations are listed in the Table~\ref{tabsh}, timings obtained using time series from the American Association of Variable Star Observers (AAVSO) are presented in the appendix~\ref{sec:online}). We built the (O-C) diagram presented in Fig.~\ref{fig:fig_ocsup} (black dots). We also plot the seasonal maxima timings published by \citet{demiguel2017} (open rectangles).

Generally, all points are scattered in a range $\pm0.4$ phase, but some group of points are outlying around the phase $0.5$ that required manual verification. The only ambiguity of cycle count may be for this group of points ($+0.5$ or $-0.5$). We claim that there are no significant changes of the superhump period from season to season, and no significant changes of the orbital period of this system on the time span of 17 years.

\begin{figure}
	\includegraphics[width=\columnwidth,trim={0.0cm 0 0.5cm 0.0cm},clip]{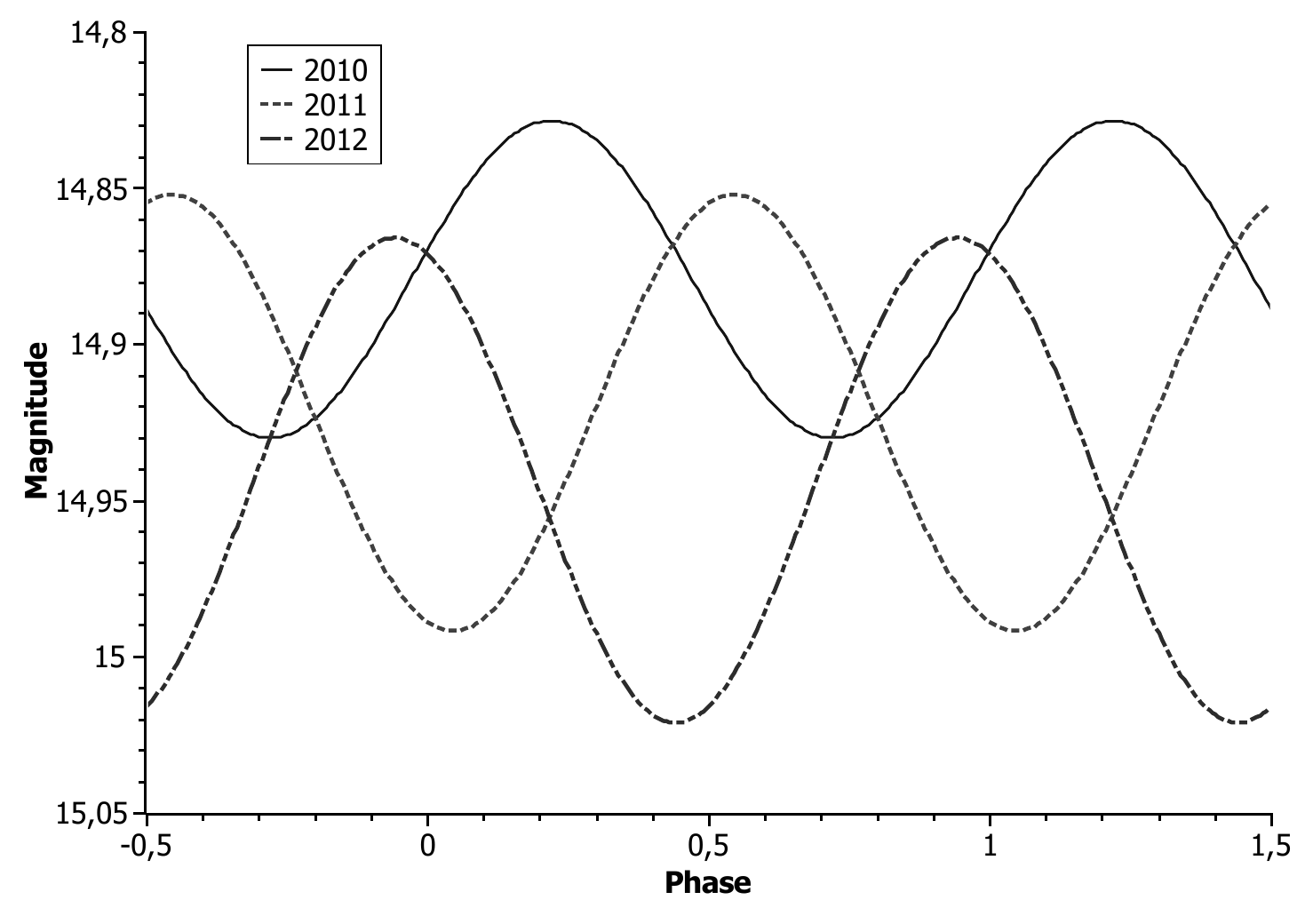}
    \caption{Sample phase curves with superhump period calculated for the data from VNT are shown.}
    \label{fig:fig_phasesh}
\end{figure}

At the same time, see Fig.~\ref{fig:fig_phasesh}, one may see changes of the superhump maxima timings from season to season. Figure~\ref{fig:fig_phasesh} displays first order trigonometric polynomial approximations of the phase curves for the selected data sets from different years, where phase curves were calculated using the same value of the period and initial epoch ($P=0.28013$, $T_0=2458130.75559$). These irregular changes of the superhump maxima timings cause the large scatter seen in Fig.~\ref{fig:fig_ocsup}.

\begin{table}
	\centering
	\caption{The seasonly values of the superhump maxima timings, semiamplitudes of the superhump modulation and phases of the O-C calculated for $P=0^d.28013$ and $T_0=2458130.75559$}
	\label{tabsh}
\begin{tabular}{cccccc}
\hline
Year & Instrument & BJD & A, mmag & $\phi$ & $\sigma\phi$ \\
\hline
2005 & LOAO & 2453671.75499 & 79(4) & 0.39 & 0.01\\
2007 & ZC600 & 2454309.20766 & 108(4) & -0.05 & 0.01\\
2008 & ZC600 & 2454686.23805 & 101(5) & -0.14 & 0.01\\
2008 & VNT & 2454718.72944 & 71(4) & -0.15 & 0.01\\
2009 & 50CM & 2455064.89775 & 86(16) & 0.59 & 0.03\\
2009 & VNT & 2455085.63884 & 67(7) & 0.63 & 0.02\\
2010 & VNT & 2455398.70925 & 51(4) & 0.22 & 0.01\\
2010 & LOAO & 2455451.69897 & 89(24) & 0.38 & 0.05\\
2011 & VNT & 2455803.86853 & 70(4) & 0.55 & 0.01\\
2011 & 50CM & 2455836.64112 & 68(10) & 0.54 & 0.02\\
2011 & LOAO & 2455848.15471 & 70(9) & 0.64 & 0.02\\
2012 & ZC600 & 2456115.21015 & 43(4) & -0.04 & 0.02\\
2012 & VNT & 2456129.21026 & 78(3) & -0.06 & 0.01\\
2012 & LOAO & 2456161.70907 & 59(13) & -0.05 & 0.04\\
2013 & LOAO & 2456446.34008 & 73(10) & 0.02 & 0.02\\
2013 & ZC600 & 2456492.26822 & 69(5) & -0.02 & 0.01\\
2013 & VNT & 2456519.14752 & 73(4) & -0.07 & 0.01\\
2014 & ZC600 & 2456823.10296 & 67(4) & -0.02 & 0.01\\
2014 & OAUJM & 2456868.74851 & 112(10) & -0.08 & 0.02\\
2014 & VNT & 2456928.74162 & 76(5) & 0.09 & 0.01\\
2015 & OAUJM & 2457164.80568 & 55(6) & -0.22 & 0.02\\
2015 & ZC600 & 2457193.65943 & 44(14) & -0.22 & 0.05\\
2015 & VNT & 2457265.23788 & 30(5) & 0.30 & 0.02\\
2016 & OAUJM & 2457573.58716 & 107(5) & 0.04 & 0.01\\
2016 & ZC600 & 2457598.81753 & 85(7) & 0.10 & 0.01\\
2016 & VNT & 2457625.09307 & 63(5) & -0.10 & 0.01\\
2016 & VNT & 2457661.55984 & 17(5) & 0.08 & 0.05\\
2017 & VNT & 2457904.70271 & 37(7) & 0.04 & 0.03\\
2017 & ZC600 & 2457940.15950 & 37(7) & -0.38 & 0.03\\
2018 & ZC600 & 2458326.94358 & 19(4) & 0.35 & 0.03\\
2018 & VNT & 2458392.30560 & 41(10) & -0.33 & 0.04\\
2019 & VNT & 2458660.16634 & 96(6) & -0.12 & 0.01\\
2019 & Meade & 2458668.86542 & 63(3) & -0.07 & 0.01\\
2019 & VNT & 2458682.59450 & 84(5) & -0.06 & 0.01\\
2019 & ZC600 & 2458723.20525 & 89(7) & -0.09 & 0.01\\
2019 & ZC600 & 2458723.76677 & 73(6) & -0.09 & 0.01\\
2020 & VNT & 2459089.33038 & 84(5) & -0.11 & 0.01\\
2021 & VNT & 2459447.95764 & 42(4) & 0.11 & 0.02\\
2022 & ZC600 & 2459805.59621 & 136(7) & -0.20 & 0.01\\
2022 & ZC600 & 2459816.80570 & 79(11) & -0.19 & 0.02\\
2022 & VNT & 2459847.37246 & 24(5) & -0.07 & 0.04\\
2023 & ZC600 & 2460139.86528 & 233(10) & 0.06 & 0.01\\
\hline
\end{tabular}
\end{table}


\section{Long term variability}
\label{sec:LongTerm}

For analysis of long-term variability, we need a large data set obtained in the same photometric system. For this purpose, we utilized the public photometry database of the All-Sky Automated Survey for Supernovae (ASAS-SN: \citet{ASAS}), that contains 194 measurements between HJD 2457009.69225 and 2458435.75585. The magnitudes are scattered in the range between 14.68 and 15.45. Any trends are not statistically significant. Data from other sources show a similar result. Period analysis of these data in the range of 10 - 1500 days show some peaks e.g. $64.36$ days or $124.35$ days and others. Phase curves calculated for such periods show variability with semiamplitudes of about $0^m.05$ that do not seem to be real. We may conclude there is no periodic long term variability, only short time non-periodic low states (see \citet{Covington2022}).


\section{Conclusions}
\label{sec:Conclusions}

From the analysis of observations gathered during 17 years of photometric monitoring of the intermediate polar RX J2133.7+5107, we determined the value of the spin-up time-scale $P_{\omega}/|\dot{P_{\omega}|} = 1.483(1)\times10^5$ yr. The observed rate of spin-up is even faster than that reported by \citet{demiguel2017} and one of the fastest of all known intermediate polars. Statistically optimal fit to the O-C diagram is the $4^{th}$ order polynomial, that argues for the variable rate of spin-up and the presence of complicated changes of the spin period, typical for this type of objects (see Fig.~\ref{fig:fig_spinocresi}). However, it also has a statistically significant coefficient $Q_4$, about $19\sigma$, and statistically better fits the data.

Variations in the white dwarf rotation rates in intermediate polars can be caused by a change of angular momentum of the white dwarf due to accretion, precession of the rotation axis of the magnetic white dwarf (see \citet{Andronov2005} and \citet{Tovmassian2007}) or the presence of a third body in the system (see \citet{2013JPhStv405}). We discovered similar variability of O-C for other intermediate polars (e.g. V2306 Cygni, \cite{breus2019wga}, FO~Aqr, \citet{breus2012fo}, MU Cam, \citet{2005JASS...22..197K}, and V405~Aur, Breus et al. (in preparation)).

We determined the value of the superhump period of $P_{sh}=0^d.28013$ and initial epoch of superhump maxima $T_0=2458130.75559$. The long-term variation of the residuals of the superhump O-C may confirm the precession of the accretion disc, that is usually associated with negative superhumps (e.g. \citet{2013JPhStv405}, \citet{Kim2009}).


\section*{Acknowledgements}
	The authors thank the referee Paul A. Mason for helpful comments resulting in the improvement of this paper.
	
	VB acknowledges financial support from the National Scholarship Programme of the Slovak Republic and Queen Jadwiga Fund of the Jagiellonian University in Krakow, Poland.
	
	This work was supported by the Slovak Research and Development Agency under the contract No. APVV-20-0148
	
	We thank RNDr. Igor Kudzej, director of Vihorlat Astronomical Observatory, Prof. Stanislaw Zola director of the Jagiellonian University's Astronomical Observatory in Krakow, Poland and Tibor Heged{\"u}s, director of Baja Astronomical Observatory, Hungary, for telescope time reservation.

	We acknowledge with thanks the variable star observations from the AAVSO International Data base (https://www.aavso.org/) contributed by observers worldwide and used in this research and personally James Jones, Shawn Dvorak, David Boyd, David Cejudo, Richard Sabo, Sjoerd Dufoer, Etienne Morelle, John Rock, William Goff, Kenneth Menzies.

\section*{Data availability}

Data from the ASAS-SN are publicly available from Sky Patrol 3. Data from AAVSO are publicly available from their Data Access page. The data obtaned by out group are available on request from the corresponding author, VB.





\begin{thebibliography}{99}
	\bibitem[\protect\citeauthoryear{Andronov \& Baklanov}{2004}]{AndBak2004} Andronov~I.\,L. \& Baklanov~A.\,V. 2004, Astron. School Reports 5, 264
	\bibitem[\protect\citeauthoryear{Andronov}{2005}]{Andronov2005} Andronov~I.\,L. 2005, ASP Conf. Ser. 334, 447
	\bibitem[\protect\citeauthoryear{Andronov et al.}{2011}]{2011JKASRXJ1803} Andronov I.~L., Kim Y., Yoon J.-N., Breus V.~V., Smecker-Hane T.~A., Chinarova L.~L., Han W., 2011, JKAS, 44, 89
	\bibitem[\protect\citeauthoryear{Andronov \& Breus}{2013}]{exhya2013} Andronov~I.\,L. \& Breus~V.\,V. 2013, Astrophysics. 56, 4, 518-530
	\bibitem[\protect\citeauthoryear{Anzolin et al.}{2009}]{Anzolin2009} Anzolin~G. \& de Martino~D. \& Falanga~M. \& Mukai~K. \& Bonnet-Bidaud~J.-M. \& Mouchet~M. \& Terada~Y. \& Ishida~M. 2009, A\&A, 501, 1047-1058.
	\bibitem[\protect\citeauthoryear{Bonnet-Bidaud et al.}{2006}]{Bonnet2006} Bonnet-Bidaud~J.\,M. \& Mouchet~M. \& de Martino~D. \& Silvotti~R. \& Motch~C. 2006, A\&A, 445, 1037-1040.
	\bibitem[\protect\citeauthoryear{Breus et al.}{2012}]{breus2012fo} Breus~V. \& Andronov~I.L. \& Hegedus~T. \& Dubovsky~P.A. \& Kudzej~I. 2012, Advances in Astronomy and Space Physics, 2, 9.
	\bibitem[\protect\citeauthoryear{Breus et al.}{2013}]{2013JPhStv405} Breus V.~V., Andronov I.~L., Dubovsky P., Kolesnikov S.~V., Zhuzhulina E.~A., Hegedus T., Beringer P., et al., 2013, JPhSt, 17, 3902.
	\bibitem[\protect\citeauthoryear{Breus et al.}{2019}]{breus2019wga} Breus~V. \& Petrik~K. \& Zola~S. 2019, MNRAS, 488, 4, 4526-4529.
	\bibitem[\protect\citeauthoryear{Covington et al.}{2022}]{Covington2022} Covington~A.\,E. \& Shaw A.\,W. Koji Mukai et al. 2022, ApJ, 928, 164.
	\bibitem[\protect\citeauthoryear{de Martino et al.}{2006}]{DeMartino2006} de Martino~D. \& Bonnet-Bidaud~J.\,M. \& Mouchet~M. \& G{\"a}nsicke~B.\,T. \& Haber~F. \& Motch~C. 2006, A\&A, 449, 1151-1160.
	\bibitem[\protect\citeauthoryear{de Miguel et al.}{2017}]{demiguel2017} de Miguel~E. et al. 2017, MNRAS, 467, 1, 428-436
	\bibitem[\protect\citeauthoryear{Katajainen et al.}{2007}]{Katajainen2007} Katajainen~S. \& Butters~O.\,W. \& Norton~A.\,J. \& Lehto~H.\,J. \& Piirola~V. 2007, A\&A, 475, 1011-1018.
	\bibitem[\protect\citeauthoryear{Kim, Andronov \& Jeon}{2004}]{KimAndMC2004} Kim~Y. \& Andronov~I.\,L. \& Jeon~Y. 2004, Journal of Astronomy and Space Sciences. 21, 3, 191-200
	\bibitem[\protect\citeauthoryear{Kim et al.}{2005}]{2005JASS...22..197K} Kim~Yong-Gi \& Andronov~I.~L. \& Park~Sung-Su \& Chinarova~L/~L. \& Baklanov~A.V. \& Jeon~ Young-Beom, 2005, Journal of Astronomy and Space Sciences, 22, 197-210
	\bibitem[\protect\citeauthoryear{Kim et al.}{2009}]{Kim2009}Kim~Y. \& Andronov~I.~L. \& Cha~S.~M. \& Chinarova~L.~L. \& Yoon~J.~N. 2009, A\&A, 496, 765-775
	\bibitem[\protect\citeauthoryear{Kochanek et al.}{2017}]{ASAS} Kochanek~C.\,S. \& Shappee~B.\,J. \& Stanek~K.\,Z. et al. 2017, PASP, 129, 104502
	\bibitem[\protect\citeauthoryear{Motch et al.}{1998}]{Motch1998} Motch~C. et al. 1998, A\&A, 132, 341-359.
	\bibitem[\protect\citeauthoryear{Motl}{2011}]{Munipack} Motl~D. C-Munipack project. 2011, http://c-munipack.sourceforge.net/
	\bibitem[\protect\citeauthoryear{Parimucha et al.}{2020}]{Caosp2020mu} Parimucha~S. \& Dubovsky~P.\,A. \& Kudzej~I. \& Breus~V. \& Petrik~K. Contrib. Astron. Obs. Skalnate Pleso 50, 618-620
	\bibitem[\protect\citeauthoryear{Patterson}{1994}]{Patterson1994} Patterson~J. 1994, PASP, 106, 209-238
	\bibitem[\protect\citeauthoryear{Savanevych et al.}{2017}]{colitec1} Savanevych~V.~E. et al. 2017, Odessa Astronomical Publications, 30, 194.
	\bibitem[\protect\citeauthoryear{Savanevych et al.}{2022}]{colitec2} Savanevych~V.~E. et al. 2022, Astronomy and Computing, 40, 100605.
	\bibitem[\protect\citeauthoryear{Thorstensen, Peters \& Skinner}{2010}]{Thorstensen2010} Thorstensen~J.\,R. \& Peters C.\,S. \& Skinner J.\,N. 2010, PASP, 122, 1285-1302.
	\bibitem[\protect\citeauthoryear{Tovmassian, Zharikov \& Neustroev}{2007}]{Tovmassian2007} Tovmassian~G.\,H. \& Zharikov~S.\,V. \& Neustroev~V.\,V. 2007, ASP Conf. Ser. 372, 541
	\bibitem[\protect\citeauthoryear{Warner}{1995}]{Warner1995} Warner, B. 1995, Cataclsymic Variables (Cambridge: Cambridge University Press)

\end{thebibliography}




\appendix
\section{Supporting Information}
\label{sec:online}
Supplementary data are available at MNRAS online.
\textbf{rxj2133-spin.txt} contains the list of spin maxima timings.
\textbf{rxj2133-sh-aavso.txt} contains lists of superhump maxima timings calculated for the data from AAVSO.
Both are tab-separated text files.



\bsp	
\label{lastpage}
\end{document}